\documentstyle[prl,aps,floats,epsf,color]{revtex}
\begin{document}
\baselineskip=12pt
\def\be{\begin{equation}}
\def\ee{\end{equation}}
\def\bea{\begin{eqnarray}}
\def\eea{\end{eqnarray}}
\def\orc{\Omega_{r_c}}
\def\om{\Omega_{\text{m}}}
\def\E{{\rm e}}
\def\bearst{\begin{eqnarray*}}
\def\eearst{\end{eqnarray*}}
\def\peleven{\parbox{11cm}}
\def\peffec{\peight{\bearst\eearst}\hfill\peleven}
\def\pspace{\peight{\bearst\eearst}\hfill}
\def\ptwelve{\parbox{12cm}}
\def\peight{\parbox{8mm}}
\twocolumn[\hsize\textwidth\columnwidth\hsize\csname@twocolumnfalse\endcsname

\title
{Consistency of $f(R)=\sqrt{R^{2}-R_{0}^2}$ Gravity with the
Cosmological Observations in Palatini Formalism}

\author{ M. Sadegh Movahed$^{1,2}$,Shant Baghram$^{3}$ and Sohrab Rahvar$^{3}$}
\address{$^{1}$Department of Physics, Shahid Beheshti University, Evin, Tehran 19839, Iran}
\address{$^{2}$ Institute for Studies in theoretical Physics and Mathematics, P.O.Box
19395-5531,Tehran, Iran}
\address{$^{3}$Department of Physics, Sharif University of
Technology, P.O.Box 11365--9161, Tehran, Iran}

\vskip 1cm

 \maketitle

\begin{abstract}
In this work we study the dynamics of universe in
$f(R)=\sqrt{R^2-R_{0}^2}$ modified gravity with Palatini formalism.
We use data from recent observations as Supernova Type Ia (SNIa)
Gold sample and Supernova Legacy Survey (SNLS) data, size of
baryonic acoustic peak from Sloan Digital Sky Survey (SDSS), the
position of the acoustic peak from the CMB observations and large
scale structure formation (LSS) from the $2$dFGRS survey to put
constraint on the parameters of the model. To check the consistency
of this action, we compare the age of old cosmological objects with
the age of universe. In the combined analysis with the all the
observations, we find the parameters of model as
$R_0=6.192_{-0.177}^{+0.167}\times H_0^2$ and
$\Omega_m=0.278_{-0.278}^{+0.273}$.
\end{abstract}
\newpage
]

\section{Introduction}
Recent Observations of Supernova type Ia (SNIa) provide the main
evidence for the accelerating expansion of the Universe
\cite{ris,permul}. Analysis of SNIa and the Cosmic Microwave
Background radiation (CMB) observations indicates that about $70\%$
of the total energy of the Universe is made by the dark energy and
the rest of it is in the form of dark matter with a few percent of
Baryonic matter \cite{bennett,peri,spe03}. The "cosmological
constant" is a possible explanation of the present dynamics of the
universe \cite{wein}. This term in Einstein field equations can be
regarded as a fluid with the equation of state of $w=-1$. However,
there are two problems with the cosmological constant, namely the
{\it fine-tuning} and the {\it cosmic coincidence}. In the framework
of quantum field theory, the vacuum expectation value is $123$ order
of magnitude larger than the observed value of $10^{-47}$ GeV$^{4}$.
The absence of a fundamental mechanism which sets the cosmological
constant to zero or to a very small value is the cosmological
constant problem. The second problem known as the cosmic
coincidence, states that why are the energy densities of dark energy
and dark matter nearly equal today?

There are various solutions for this problem as the decaying
cosmological constant models. A non-dissipative minimally coupled
scalar field, so-called Quintessence can play the role of this time
varying cosmological constant \cite{wet88,amen,peb88}. The ratio of
energy density of this field to the matter density in this model
increases by the expansion of the universe and after a while dark
energy becomes the dominant term of the energy-momentum tensor. One
of the features of this model is the variation of equation of state
during the expansion of the universe. Various Quintessence models
like k-essence \cite{arm00}, tachyonic matter \cite{pad03}, Phantom
\cite{cal02,cal03} and Chaplygin gas \cite{kam01} provide various
equations of states for the dark energy
{\cite{cal03,arb05,wan00,per99,pag03,dor01,dor02,dor04}.

Another approach dealing with this problem is using the modified
gravity by changing the Einstein-Hilbert action. Recently a great
attention has been devoted to this era because of the prediction of
early and late time accelerations in these models \cite{modi}. While
it seems that the modified gravity and dark energy models are
completely different approaches to explain cosmic acceleration, it
is possible to unify them in one formalism \cite{sob07}.

 In this work we obtain the dynamics of the
modified gravity $f(R)=\sqrt{R^{2}-R_{0}^{2}}$ in the Palatini
formalism \cite{bagh} and use the cosmological observations as SNIa,
SDSS acoustic peck in CMB and structure formation to put constraint
on the parameter of the action. In Sec. \ref{section2} we derive the
dynamics of Hubble parameter and scale factor in Palatini formalism
for general case of $f(R)$ gravity. In Sec. \ref{sec3} using action
$f(R)=\sqrt{R^{2}-R_{0}^{2}}$ we obtain the dynamics of universe. In
Sec. \ref{sec4} we use the observations from the evolution of
background as SNIa, CMB and baryonic acoustic oscillation to
constrain the parameters of the model. Sec. \ref{sec5} studies
constraints from the large scale structure formation and in Sec.
\ref{sec6} we compare the age of universe from the model with the
age of old cosmological objects. The conclusion is presented in Sec.
\ref{conc}.

\section{modified gravity models in Palatini}
\label{section2} An alternative approach dealing with the
acceleration problem of the Universe is changing the gravity law
through the modification of action of gravity by means of using
$f(R)$ instead of the Einstein-Hilbert action. For an arbitrary
action of the gravity there are two main approaches to extract the
field equations. The first one is the so-called ''metric formalism''
in which the variation of action is performed with respect to the
metric. In the second approach, ''Palatini formalism'', the
connection and metric are considered independent of each other and
we have to do variation for those two parameters independently.
General form of the action in palatini formalism is :
\begin{equation}
S[f;g,\hat{\Gamma},\Psi_{m}]=-\frac{1}{2\kappa}\int{d^{4}x\sqrt{-g}f(R)+S_{m}[g_{\mu\nu},\Psi_{m}]},
\end{equation}
where $\kappa=8\pi{G}$ and $S_{m}[g_{\mu\nu},\Psi_{m}]$ is the
matter action which depends only on metric $g_{\mu\nu}$ and on the
matter fields $\Psi_{m}$.
$R=R(g,\hat{\Gamma})=g^{\mu\nu}{R_{\mu\nu}}(\hat{\Gamma})$ is the
generalized Ricci scalar and $R_{\mu\nu}$ is the Ricci tensor of the
affine connection which is independent of the metric. Varying the
action with respect to the metric results in:
\begin{equation}
f'(R)R_{\mu\nu}(\hat{\Gamma})-\frac{1}{2}f(R)g_{\mu\nu}=\kappa
T_{\mu\nu}, \label{field}
\end{equation}
where prime is the differential with respect to Ricci scalar and
$T_{\mu\nu}$ is the energy momentum tensor
\begin{equation}
T_{\mu\nu}=\frac{-2}{\sqrt{-g}}\frac{\delta S_{m}}{\delta
g^{\mu\nu}}.
\end{equation}
Varying the action with respect to the  connection and after
contraction gives us the equation that determines the  generalized
connection as:
\begin {equation}
\hat{\nabla_{\alpha}}[f'(R)\sqrt{-g}g^{\mu\nu}]=0,
\end {equation}
where $\hat{\nabla}$ is the covariant derivative with respect to the
affine connection. We can see that the connections are the
Christoffel symbols of the new metric $h_{\mu\nu}$ where it is
conformally related to the original one via the
$h_{\mu\nu}=f'(R)g_{\mu\nu}$.

Equation (\ref{field}) shows that in contrast to the metric
variation approach, the field equations are second order in this
formalism which seems more intuitive to have second order equations
instead of fourth order. On the other hand forth order differential
equations have the instability  problem \cite{nest07}. We use FRW
metric for universes as follows
\begin{equation}
ds^{2}=-dt^{2}+a(t)^{2}\delta_{ij}dx^{i}dx^{j}
\end{equation}
and assume universe is filled with perfect fluid with the
energy-momentum tensor of $T^{\nu}_{\mu}=diag(-\rho,p,p,p)$, taking
the trace of equation (\ref{field}) gives,
\begin{equation}
Rf'(R)-2f(R)=\kappa T, \label{trace}
\end{equation}
where $T=g^{\mu\nu}T_{\mu\nu}=-\rho+3p$. Using Generalized Einstein
equation (\ref{field}) and FRW metric we obtain generalized FRW
equation as \cite{all04a}:
\begin{equation}
(H+\frac{1}{2}\frac{\dot{f'}}{f'})^{2}=\frac{1}{6}\frac{\kappa(\rho+3p)}{f'}-\frac{1}{6}\frac{f}{f'}.
\label{hp}
\end{equation}
For the cosmic fluid with the equation of state of $p=p(\rho)$,
using the continuity equation and the trace of field equation we can
express the time derivative of Ricci scalar as:
\begin{equation}
\dot{R}=+3H\frac{(1-3p')(\rho+p)}{Rf''-f'(R)}.
\end{equation}
Using equation (\ref{trace}) we obtain the density of matter in
terms of Ricci scalar as:
\begin{equation} \label{den}
\kappa\rho=\frac{2f-Rf'}{1-3\omega},
\end{equation}
where $w=p/\rho$. Substituting (\ref{den}) in (\ref{hp}) we obtain
the dynamics of universe in terms of Ricci scalar:
\begin{equation}
H^{2}=\frac{1}{6(1-3\omega)f'}\frac{3(1+\omega)f-(1+3\omega)Rf'}{\left[1+\frac{3}{2}(1+\omega)\frac{f'(Rf'-2f)}{f'(Rf''-f')}\right]^{2}}.
\label{hpala}
\end{equation}
On the other hand using equation ({\ref{trace}) and continuity
equation, the scale factor can be obtained in terms of Ricci scalar
\begin{equation}
a=\left[\frac{1}{\kappa\rho_{0}(1-3\omega)}(2f-Rf')\right]^{-\frac{1}{3(1+\omega)}},
\label{apala}
\end{equation}
where $\rho_{0}$ is the energy density at the present time and we
set $a_{0}=1$. Now for a modified gravity action, omitting Ricci
scalar in favor of the scale factor between equations (\ref{hpala})
and (\ref{apala}) we can obtain the dynamics of universe (i.e.
$H=H(a)$). For the matter dominant epoch $\omega=0$, these equations
reduce to:
\begin{eqnarray} \label{Hub}
H^{2}&=&\frac{1}{6f'}\frac{3f-R f'}{\left[1+\frac{3}{2}\frac{f''(R f'-2f)}{f'(R f''-f{'})}\right]^{2}}\\
a&=&\left[\frac{1}{\kappa\rho_{0}}(2f-R f')\right]^{-\frac{1}{3}}
\end{eqnarray}
\section{Modified gravity with the action of $f(R)=\sqrt{R^{2}-R_{0}^{2}}$}
\label{sec3} In this section our aim is to apply the action of
$f(R)=\sqrt{R^{2}-R_{0}^{2}}$ in equations (\ref{hpala}) and
(\ref{apala}) to obtain the dynamics of universe. For simplicity we
use dimensionless parameters in our calculation defined by
\begin{equation}
f(R)=H_{0}^2\sqrt{X^{2}-X_{0}^{2}}
\label{fx}
\end{equation}
in which $H_{0}$ is the Hubble parameter at the present time and
$X\equiv\frac{R}{H_{0}^2}$ and $X_{0}\equiv\frac{R_{0}}{H_{0}^2}$.
For convenience we can write the action and its derivatives as:
\begin{equation}
f(R)=H_{0}^{2}F(X)
\end{equation}
\begin{equation}
f'(R)=F'(X)
\end{equation}
\begin{equation}
f''(R)=\frac{F''(X)}{H^2_{0}}
\end{equation}
where derivative at the right hand side of equations is with respect
to $X$ (i.e. $'=\frac{d}{dX}$). We can write (\ref{Hub}) with new
parameter of $X$ as:
\begin{equation}
{\cal{H}}(X)=\frac{1}{6F'}\frac{3F-XF'}{(1+\frac{3}{2}\frac{F''(XF'-2F)}{F'(XF''-F')})^{2}},
\label{hcal}
\end{equation}
where ${\cal{H}}(X)= {H^{2}}/{H_{0}^2}$ and $H_0$ is the Hubble
parameter at the present time. Using the definition of
$\Omega_{m}(X)\equiv\kappa\rho_{m}/(3H^{2})$ from equation
(\ref{den}) we can obtain $\Omega_m(X)$ in terms $X$ as:
\begin{equation}\label{matter}
\Omega_{m}(X)=\frac{2F-XF'}{3}. \label{xf}
\end{equation}
We use $\Omega_m(X) = \Omega_m a^{-3}$ (where $\Omega_{m}\equiv
\Omega_{m}^{(0)}$) and substitute F in terms of $X$ from (\ref{fx})
to obtain scale factor in terms of $X$ as:
\begin{equation}\label{apala1}
a=\left[\frac{1}{3\Omega_{m}}(\frac{X^{2}-2X_{0}^{2}}{\sqrt{X^{2}-X_{0}^2}})\right]^{-1/3},
\end{equation}
where to have positive scale factor, $X$ should change in the range
of $ X\geq\sqrt{2} X_{0}$. It should be noted that this model has
only one free parameter of $X_0$. $X_p$ represents the value of $X$
at the present time and from (\ref{hcal}) we can find $X_p$ in terms
of $X_0$. On the other hand substituting $X_p=g(X_0)$ in (\ref{xf})
we will have direct relation between $X_0$ and $\Omega_m$. So
knowing $X_0$ from observation one can calculate also $\Omega_m$.

In the next section we will compare observations with the dynamics
provided by this model in the matter dominant epoch. However we can
see how the universe expands at the radiation dominant epoch. We set
$p=1/3\rho$ which implies traceless energy momentum tensor and using
our proposed action in equation (\ref{trace}), we get a constant
Ricci scalar of $R = \sqrt{2} R_0$ for this area. Substituting this
constant curvature in equation (\ref{hp}) we have:
\begin{equation}
H^2 = \frac{\kappa}{3\sqrt{2}}\rho - \frac{1}{6\sqrt{2}}R_0.
\end{equation}
Since $R_0$ is in the order of $H_0^2$ (see \cite{bagh}), for the
early universe we can ignore second term at the right hand side of
this equation which results the scale factor to increase as $a
\propto t^{1/2}$.

\section{Observational Constraints From the Background Evolution}
\label{sec4}
 In this section we compare the new SNIa Gold sample and supernova Legacy Survey data,
the location of baryonic acoustic oscillation peak from the SDSS and
the location of acoustic peak from the CMB observation to constrain
the parameters of the model. We choose various priors applied in the
this analysis as shown in Table \ref{tab1}
\begin{table}
\begin{center}
\caption{\label{tab1} Different priors on the parameter space, used
in the likelihood analysis.}
\medskip
\begin{tabular}{|c|c|c|}
  {\rm Parameter}& Prior & \\  \hline
  $\Omega_{tot}$ & $1.00$ & {\rm Fixed}\\\hline
 $\Omega_bh^2$&$0.020\pm0.005$&{\rm Top hat (BBN)\cite{bbn}}\\\hline
 $h$&$-$&{\rm Free \cite{hst,zang}}\\\hline
 $w$&$0$&{\rm Fixed}\\
 \end{tabular}
\end{center}
\end{table}

\subsection{Examining the Model by Supernova Type Ia: Gold Sample}
\label{sn} The Supernova Type Ia experiments provided the main
evidence for the present acceleration of the universe. Since 1995
two teams of the {\it High-Z Supernova Search} and the {\it
Supernova Cosmology Project} have discovered several type Ia
supernovas at the high redshifts \cite{per99,Schmidt}.  Riess et
al.(2004) announced the discovery of $16$ type Ia supernova with the
Hubble Space Telescope. This new sample includes $6$ of the $7$ most
distant ($z> 1.25$) type Ia supernovas. They determined the
luminosity distance of these supernovas and with the previously
reported algorithms, obtained a uniform $157$ Gold sample of type Ia
supernovas \cite{R04,Tonry,bar04}. Recently a new data set of Gold
sample with smaller systematic error containing $156$ Supernova Ia
has been released \cite{new}. In this work we use this data set as
new Gold sample SNIa.

More recently, the SNLS collaboration released the first year data
of its planned five-year Supernova Legacy Survey\cite{astier05}. An
important aspect to be emphasized on the SNLS data is that they seem
to be in a better agreement with WMAP results than the Gold sample
\cite{pad06}.

We calculate the apparent magnitude from the $f(R)$ modified gravity
and compare with new SNIa Gold sample and SNLS data set. The
supernova measured apparent magnitude $m$ includes reddening, K
correction, etc, which here all these effect have been removed. The
apparent magnitude is related to the (dimensionless) luminosity
distance, $D_L$, of an object at redshift $z$ through:
\begin{equation}
m={\mathcal{M}}+5\log{D_{L}(z;X_{0})}, \label{m}
\end{equation} where
\begin{eqnarray}
\label{luminosity} D_L (z;X_{0}) &=&(1+z) \int_0^z\, {dz'H_0\over
H(z')}
\end{eqnarray}
Also
\begin{eqnarray}
\label{m1}\mathcal{M} &=& M+5\log{\left(\frac{c/H_0}{1\quad
Mpc}\right)}+25.
\end{eqnarray}
where $M$ is the absolute magnitude. The distance modulus, $\mu$, is
defined as:

\begin{eqnarray}
\mu\equiv m-M&=&5\log{D_{L}(z;X_{0})}\nonumber\\&&\quad
+5\log{\left(\frac{c/H_0}{1\quad Mpc}\right)}+25, \label{eq:mMr}
\end{eqnarray}
or

\begin{eqnarray}
\mu&=&5\log{D_{L}(z;X_{0})}+\bar{M}.
\end{eqnarray}
\begin{figure}[t]
\epsfxsize=9.5truecm
\begin{center}
\epsfbox{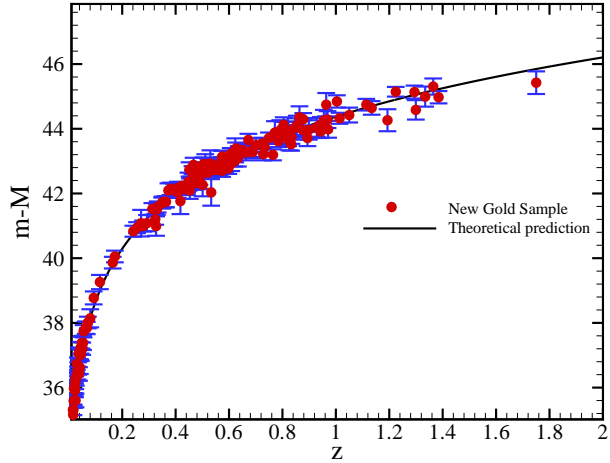} \narrowtext \caption{Distance modulus of the SNIa
new Gold sample in terms of redshift. Solid line shows the best fit
values with the corresponding parameters of $h=0.63$,
$\Omega_m=0.276_{-0.240}^{+0.376}$, $X_0=6.207^{+0.230}_{-0.147}$ in
$1 \sigma$ level of confidence with $\chi^2_{min}/N_{d.o.f} =0.912$
for $f(R)$ model.} \label{1m}
\end{center}
\end{figure}
\begin{figure}[t]
\epsfxsize=9.5truecm
\begin{center}
\epsfbox{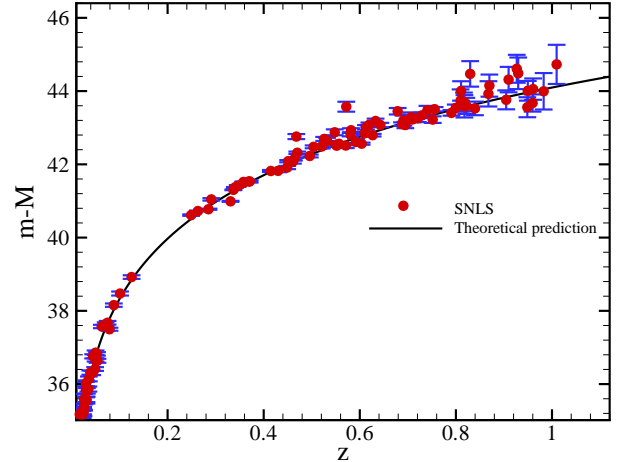} \narrowtext \caption{Distance modulus of the SNLS
supernova data in terms of redshift. Solid line shows the best fit
values with the corresponding parameters of $h=0.70$,
$\Omega_m=0.246_{-0.164}^{+0.164}$ and $X_0=6.484^{+0.099}_{-0.099}$
in $1 \sigma$ level of confidence with $\chi^2_{min}/N_{d.o.f}
=11.766$ for $f(R)$ model.} \label{2m}
\end{center}
\end{figure}
To compare the theoretical results with the observational data, we
calculate theoretical distance modulus. Distance modulus can be
written in terms of new parameter $X$ as:
\begin{equation}
D_{L}=\frac{1}{3}\frac{(XF'-2F)^{(\frac{1}{3})}}{(3\Omega_m)^{\frac{2}{3}}}\int_{X_{p}}^{X}\frac{XF''-F'}{(XF^{'}-2F)^{\frac{2}{3}}}\frac{dX}{{\cal
H }(X)}.
\end{equation}

To put constraint on the model parameter, the first step is to
compute the quality of the fitting through the least squared fitting
quantity $\chi^2$ defined by:
\begin{eqnarray}\label{chi_sn}
\chi^2(\bar{M},X_{0})&=&\sum_{i}\frac{[\mu_{obs}(z_i)-\mu_{th}(z_i;X_0,\bar{M})]^2}{\sigma_i^2},\nonumber\\
\end{eqnarray}
where $\sigma_i$ is all of observational uncertainty in the distance
modulus. To constrain the parameters of model, we use the Likelihood
statistical analysis:
\begin{eqnarray}
{\cal L}(\bar{M},X_{0})={\mathcal{N}}e^{-\chi^2(\bar{M},X_{0})/2},
\end{eqnarray}
where ${\mathcal{N}}$ is a normalization factor. The parameter
$\bar{M}$ is a nuisance parameter and should be marginalized
(integrated out) leading to a new $\bar{\chi}^{2}$ defined as:
\begin{eqnarray}\label{mar2}
\bar{\chi}^2=-2\ln\int_{-\infty}^{+\infty}e^{-\chi^2/2}d\bar{M}.
\end{eqnarray}
 Using equations (\ref{chi_sn}) and (\ref{mar2}), we find:
\begin{eqnarray}\label{mar3}
\bar{\chi}^2(X_{0})&=&\chi^2(\bar{M}=0,X_{0})-\frac{B(X_{0})^2}{C}+\ln(C/2\pi),
\end{eqnarray}
where
\begin{eqnarray}\label{mar4}
B(X_{0})=\sum_{i}\frac{[\mu_{obs}(z_i)-\mu_{th}(z_i;X_{0},\bar{M}=0)]}{\sigma_i^2},
\end{eqnarray}
and
\begin{eqnarray}\label{mar5}
C=\sum_{i}\frac{1}{\sigma_i^2}.
\end{eqnarray}
Equivalent to marginalization is the minimization with respect to
$\bar{M}$. One can show that $\bar\chi^2$ can be expanded in terms
of $\bar{M}$ as \cite{Nesseris04}:
\begin{eqnarray}\label{mar6}
\chi^2_{\rm
SNIa}(X_{0})=\chi^2(\bar{M}=0,X_{0})-2\bar{M}B+\bar{M}^2C,
\end{eqnarray}
which has a minimum for $\bar{M}=B/C$:
\begin{eqnarray}\label{mar7}
\chi^2_{\rm
SNIa}(X_{0})=\chi^2(\bar{M}=0,X_{0})-\frac{B(X_{0})^2}{C}.
\end{eqnarray}

Using equation (\ref{mar7}) we can find the best fit values of model
parameters, minimizing $\chi^2_{\rm SNIa}(X_{0})$. Using new Gold
sample SNIa, the best fit values for the free parameter of the model
is $X_0=6.207_{-0.147}^{+0.230}$ states
$\Omega_m=0.276_{-0.240}^{+0.376}$, with
$\chi^2_{min}/N_{d.o.f}=0.912$ at $1 \sigma$ level of confidence.
The corresponding value for the Hubble parameter at the minimized
$\chi^2$ is $h=0.64$ and since we have already marginalized over
this parameter we do not assign an error bar for it. The best fit
values for the parameters of model by using SNLS supernova data is
$X_0=6.484_{-0.099}^{+0.099}$ indicates
$\Omega_m=0.246_{-0.164}^{+0.164}$ with
$\chi^2_{min}/N_{d.o.f}=11.766$ at $1 \sigma$ level of confidence.
Figures \ref{1m} and \ref{2m} show the comparison of the theoretical
prediction of distance modulus by using the best fit value of $X_0$
and observational values from new Gold sample and SNLS supernova Ia,
respectively. In figures \ref{like1}, \ref{like2}, relative
likelihood for $X_0$ are indicated. We report the best value of
$X_0$ at 1 and 2-$\sigma$ confidence level and other derived
parameters in table \ref{tab2}.

\subsection{CMBR Shift parameter}
Before last scattering, the photons and baryons are tightly coupled
by Compton scattering and behave as a fluid. The oscillations of
this fluid, occurring as a result of the balance between the
gravitational interactions and the photon pressure, lead to the
familiar spectrum of peaks and troughs in the averaged temperature
anisotropy spectrum which we measure today. The odd peaks correspond
to maximum compression of the fluid, the even ones to rarefaction
\cite{hu96}. In an idealized model of the fluid, there is an
analytic relation for the location of the m-th peak: $l_m \approx
ml_A$ \cite{Hu95,hu00} where $l_A$ is the acoustic scale which may
be calculated analytically and depends on both pre- and
post-recombination physics as well as the geometry of the universe.
The acoustic scale corresponds to the Jeans length of photon-baryon
structures at the last scattering surface some $\sim 379$ Kyr after
the Big Bang \cite{spe03}. The apparent angular size of acoustic
peak can be obtained by dividing the comoving size of sound horizon
at the decoupling epoch $r_s(z_{dec})$ by the comoving distance of
observer to the last scattering surface $r(z_{dec})$:
\begin{equation}
\theta_A =\frac{\pi}{l_A}\equiv {{r_s(z_{dec})}\over r(z_{dec}) }.
\label{eq:theta_s}
\end{equation}
The size of sound horizon at the numerator of equation
(\ref{eq:theta_s}) corresponds to the distance that a perturbation
of pressure can travel from the beginning of universe up to the last
scattering surface and is given by:
 \begin{equation}
 r_{s}(z_{dec})= \int_{z_{dec}}^ {\infty} {v_s(z')dz'\over H(z')/H_0} \label{sh}
 \end{equation}
where $v_s(z)^{-2}=3 + 9/4\times\rho_b(z)/\rho_{rad}(z)$ is the
sound velocity in the unit of speed of light from the big bang up to
the last scattering surface \cite{dor01,Hu95} and the redshift of
the last scattering surface, $z_{dec}$, is given by \cite{Hu95}:
\begin{eqnarray}\label{dec}
z_{dec} &=& 1048\left[ 1 + 0.00124(\omega_b)^{-0.738}\right]\left[
1+g_1(\omega_m)^{g_2}\right],\nonumber \\
g_1 &=& 0.0783(\omega_b)^{-0.238}\left[1+39.5(\omega_b)^{0.763}\right]^{-1},\nonumber \\
g_2 &=& 0.560\left[1 + 21.1(\omega_b)^{1.81}\right]^{-1},
\end{eqnarray}
where $\omega_m\equiv\Omega_mh^2$ and  $\omega_b\equiv\Omega_bh^2$.
Changing the parameters of the model can change the size of apparent
acoustic peak and subsequently the position of $l_A\equiv
\pi/\theta_A$ in the power spectrum of temperature fluctuations on
CMB. The simple relation $l_m\approx ml_A$ however does not hold
very well for the first peak although it is better for higher peaks
[2]. Driving effects from the decay of the gravitational potential
as well as contributions from the Doppler shift of the oscillating
fluid introduce a shift in the spectrum. A good parameterization for
the location of the peaks and troughs is given by
\cite{hu00,doran01}
\begin{equation}\label{phase shift}
l_m=l_A(m-\phi_m)
\end{equation}
where $\phi_m$ is phase shift determined predominantly by
pre-recombination physics, and are independent of the geometry of
the Universe. Instead of the peak locations of power spectrum of
CMB, one can use another model independent parameter which is
so-called shift parameter ${\cal R}$ as:
\begin{equation}
{\cal R}\propto\frac{l_1^{flat}}{l_1}
\end{equation}
where $l_1^{flat}$ corresponds to flat pure-CDM model with
$\Omega_m=1.0$ and the same $\omega_m$ and $\omega_b$ as the
original model. The location of  first acoustic peak can be
determined in model by equation (\ref{phase shift}) with
$\phi_1(\omega_m,\omega_b)\simeq 0.27$ \cite{hu00,doran01}. It is
easily shown that shift parameter is as follows \cite{bond97}:
\begin{equation}
\label{shift} {\cal R}=
\sqrt{\Omega_m}\frac{D_L(z_{dec};X_0)}{(1+z_{dec})}
\end{equation}
In writing the shift parameter in this form we have implicitly
assumed that photons follow geodesics determined by the Levi-Civita
connection. Furthermore, in order to use the shift parameter, the
evolution of the universe is considered in such a way that at the
decoupling we have standard matter dominated behavior. Since we do
not have an explicit expression for the Hubble parameters in terms
of redshift, it is useful to rewrite the shift parameter in terms of
dimensionless parameter $X$ as follows:

\begin{eqnarray}
&&{\cal{R}}=\sqrt{\Omega_{m}H_{0}^{2}}\int_{0}^{z_{dec}}\frac{dz}{H(z)}\nonumber\\
&&=\sqrt{\Omega_{m}{H_{0}}^{2}}\int_{R_{dec}}^{R_{p}}\frac{a'(R)}{a^{2}(R)}\frac{dR}{H(R)}\nonumber\\
&&=\frac{\Omega_{m}^{\frac{1}{6}}}{3^{\frac{4}{3}}}\int_{X_{p}}^{X_{dec}}
    \frac{F'-XF''}{(2F-XF^{'})^{\frac{2}{3}}}\frac{dX}{{\cal{H}}(X)}
\end{eqnarray}
The observational results of CMB experiments correspond to a shift
parameter of ${\cal R}=1.716\pm0.062$ (given by WMAP, CBI, ACBAR)
\cite{spe03,pearson03}. One of the advantages of using the parameter
${\cal{R}}$ is that it is independent of Hubble constant. In order
to put constraint on the model from CMB, we compare the observed
shift parameter with that of model using likelihood statistic as
\cite{bond97}:
\begin{equation}
{\cal{L}}\sim e^{-\chi_{\rm CMB}^2/2}
\end{equation}
where \begin{equation}\label{chi_cmb} \chi_{\rm
CMB}^2=\frac{\left[{\cal R}_{{\rm obs}}-{\cal R}_{{\rm
the}}\right]^2}{\sigma_{\rm CMB}^2}
\end{equation}

\begin{figure}[t]
\begin{center}
\epsfxsize=9.5truecm\epsfbox{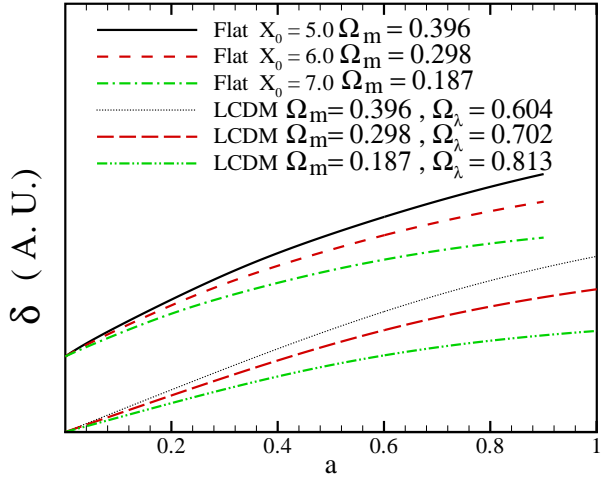} \narrowtext
\caption{Evolution of density contrast in the $f(R)$ gravity model
versus redshift for different values of $X_0$. For comparison we
plot $\delta$ for $\Lambda$CDM with the same value of
$\Omega_m$.(A.U. stands for arbitery unit)} \label{ev}
\end{center}
\end{figure}

\subsection{Baryon Acoustic Oscillations}

The large scale correlation function measured from $46,748$ {\it
Luminous Red Galaxies} (LRG) spectroscopic sample of the SDSS (Sloan
Digital Sky Survey) includes a clear peak at about  $100$ Mpc
$h^{-1}$ \cite{eisenstein05}. This peak was identified with the
expanding spherical wave of baryonic perturbations originating from
acoustic oscillations at recombination. The comoving scale of this
shell at recombination is about 150Mpc in radius. A dimensionless
and independent of $H_0$ of this observation is:
\begin{equation} \label{lss1}
{\cal A} = \sqrt{\Omega_m}\left[\frac{H_0D_L^2(z_{\rm
sdss};X_0)}{H(z_{\rm sdss};X_0)z_{\rm sdss}^2(1+z_{\rm
sdss})^2}\right]^{1/3}.
\end{equation}
or
\begin{equation}
{\cal
A}=\sqrt{\Omega_{m}}{\cal{H}}(X)^{-\frac{1}{3}}{\left[\frac{1}{z_{\rm
sdss}}\int_{0}^{z_{\rm
sdss}}\frac{dz}{{\cal{H}}(X)}\right]}^{\frac{2}{3}}
\end{equation}
we can write the above dimensionless quantity in term of our model
parameter as:

\begin{eqnarray}
A&=&\sqrt{\Omega_{m}}{\cal{H}}(X)^{-\frac{1}{3}}\nonumber\\
&&\times \left[\frac{(3\Omega_m)^{-\frac{1}{3}}}{3z_{\rm
sdss}}\int_{X_{p}}^{X_{\rm
sdss}}\frac{(F'-XF'')dX}{{\cal{H}}(X)(2F-XF')^{2/3}}\right]^{\frac{2}{3}}
\end{eqnarray}
We can put the robust constraint on the $f(R)$ modified gravity
model using the value of ${\cal A}=0.469\pm0.017$ from the LRG
observation at $z_{\rm sdss} = 0.35$ \cite{eisenstein05}. This
observation permits the addition of one more term in the $\chi^2$ of
equations (\ref{mar7}) and (\ref{chi_cmb}) to be minimized with
respect to $H(z)$ model parameters. This term is:
\begin{equation}
\chi_{\rm SDSS}^2=\frac{\left[{\cal A}_{\rm obs}-{\cal A}_{\rm
the}\right]^2}{\sigma_{\rm SDSS}^2}.
\end{equation}
This is the third observational constraint for our analysis.

\begin{figure}[t]
\begin{center}
\epsfxsize=9.5truecm\epsfbox{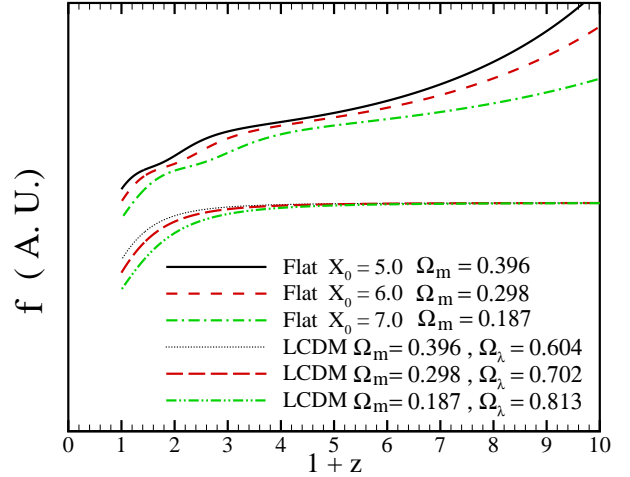} \narrowtext \caption{Growth
index versus redshift for different values of $X_0$. To make sense
we plot $f$ for $\Lambda$CDM with the same $\Omega_m$.} \label{gi}
\end{center}
\end{figure}

\subsection{Combined analysis: SNIa$+$CMB$+$SDSS} \label{cmb}
In this section we combine SNIa data (from SNIa new Gold sample and
SNLS), CMB data from the WMAP with recently observed baryonic peak
from the SDSS to constrain the parameter of modified gravity model
by minimizing the combined $\chi^2 = \chi^2_{\rm
{SNIa}}+\chi^2_{{\rm CMB}}+\chi^2_{{\rm SDSS}}$ \cite{b03}.

The best values of the model parameters from the fitting with the
corresponding error bars from the likelihood function marginalizing
over the Hubble parameter in the multidimensional parameter space
are: $X_0=6.169_{-0.184}^{+0.170}$ corresponds to
$\Omega_m=0.281_{-0.281}^{+0.277}$at $1\sigma$ confidence level with
$\chi^2_{min}/N_{d.o.f}=0.902$. The Hubble parameter corresponding
to the minimum value of $\chi^2$ is $h=0.63$. Here we obtain an age
of $14.56_{-0.28}^{+0.29}$ Gyr for the universe. Using the SNLS
data, the best fit values of model parameter is:
$X_0=6.428_{-0.092}^{+0.00}$ states
$\Omega_m=0.252_{-0.152}^{+0.148}$ at $1\sigma$ confidence level
with $\chi^2_{min}/N_{d.o.f}=11.582$. Table \ref{tab2} indicates the
best fit values for the cosmological parameters with one and two
$\sigma$ level of confidence. Relative likelihood analysis for $X_0$
using CMB and SDSS observations are shown in figures \ref{like1} and
\ref{like2}.

\begin{figure}[t]
\epsfxsize=9.5truecm\epsfbox{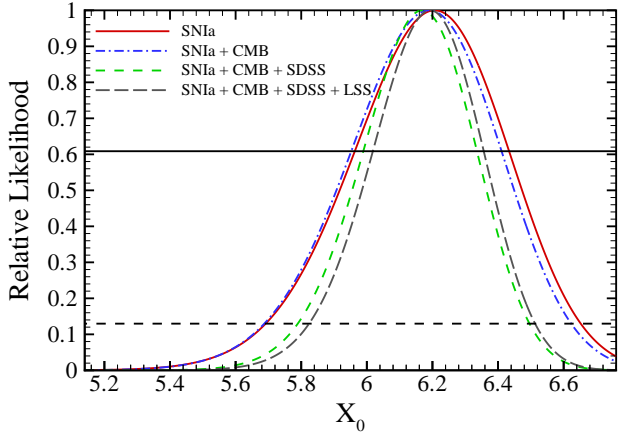} \narrowtext
\caption{Marginalized likelihood functions of $f(R)$ modified
gravity model free parameter, $X_0$. The solid line corresponds to
the likelihood function of fitting the model with SNIa data (new
Gold sample), the dashdot line with the joint SNIa$+$CMB$+$SDSS data
and dashed line corresponds to SNIa$+$CMB$+$SDSS$+$LSS. The
intersections of the curves with the horizontal solid and dashed
lines give the bounds with $1\sigma$ and $2\sigma$ level of
confidence respectively.} \label{like1}
\end{figure}
\begin{figure}[t]
\epsfxsize=9.5truecm\epsfbox{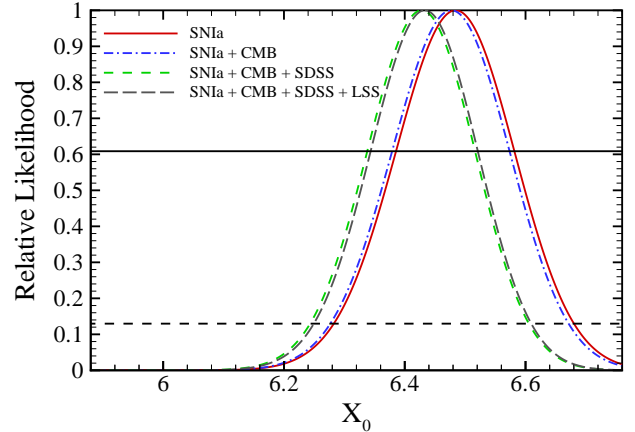} \narrowtext
\caption{Marginalized likelihood functions of $f(R)$ modified
gravity model free parameter, $X_0$. The solid line corresponds to
the likelihood function of fitting the model with SNIa data (SNLS),
the dashdot line with the joint SNIa$+$CMB$+$SDSS data and dashed
line corresponds to SNIa$+$CMB$+$SDSS$+$LSS. The intersections of
the curves with the horizontal solid and dashed lines give the
bounds with $1\sigma$ and $2\sigma$ level of confidence
respectively.} \label{like2}
\end{figure}


\section{Constraints by Large Scale structure }
\label{sec5}
So far we have only considered observational results
related to the background evolution. In this section using the
linear approximation of structure formation we obtain the growth
index of structures and compare it with the result of observations
by the $2$-degree Field Galaxy Redshift Survey ($2$dFGRS). The
continuity and modified Poisson equations for the density contrast
$\delta=\delta \rho/\bar{\rho}$ in the cosmic fluid provide the
evolution of density contrast in the linear approximation (i.e.
$\delta\ll 1 $)\cite{P93,B04} as:
\begin{equation}
\ddot{\delta}+2\frac{\dot{a}}{a}\dot{\delta} - \left[ {v_s}^2
\nabla^2 +4\pi G \rho\right] \delta=0, \label{eq2}
\end{equation}
the dot denotes the derivative with respect to time. The effect of
modified gravity in the evolution of the structures in this equation
enters through its influence on the expansion rate.

The validity of this linear Newtonian approach is restricted to
perturbations on the sub-horizon scales but large enough where
structure formation is still in the linear regime
\cite{P93,B04,maartens2}. For the perturbations larger than the
Jeans length, $ \lambda_J= \pi^{1/2} v_s / \sqrt{G \rho} $, equation
(\ref{eq2}) for cold dark matter (CDM) reduces to:
\begin{equation} \ddot{\delta}+2 \frac{\dot{a}}{a} \dot{\delta}-4\pi G \rho\delta=0. \label{eq3}
\end{equation}
The equation for the evolution of density contrast can be
rewritten in terms of the scale factor as:
\begin{equation}
\frac{d^2\delta}{da^2}+\frac{d\delta}{da}\left[\frac{\ddot{a}}{\dot{a}^2}+\frac{2H}{\dot{a}}\right]-
\frac{3H_0^2}{2\dot{a}^2a^3}\Omega_{m}\delta=0. \label{eq31}
\end{equation}
In order to use the constraint from large scale structure we should
rewrite above equation in terms of $X$. So we have
\begin{eqnarray}\label{fu1}
\dot{a}&=& a H(X),\nonumber\\
\ddot{a}&=&a H^2(X)+a^2H(X)H'(X)\frac{dX}{da}.
\end{eqnarray}
Using equation (\ref{fu1}), equation (\ref{eq31}) becomes:

\begin{eqnarray}\label{fu2}
\frac{d^2\delta}{da^2}+\frac{d\delta}{da}\left[\frac{3}{a}+\frac{{\cal
H}'(X)}{{\cal
H}(X)}\frac{dX}{da}\right]-\frac{3\Omega_m}{2{\cal{H}}^2(X)a^5}\delta=0
\end{eqnarray}

Numerical solution of equation (\ref{fu2}) in FRW universe for the
two cases of $f(R)$ gravity and $\Lambda$CDM model with the same
matter density content is compared in Figure (\ref{ev}).

In the linear perturbation theory, the peculiar velocity field
$\bf{v}$ is determined by the density contrast \cite{P93,P80} as:
\begin{equation} {\bf v} ({\bf x})= H_0 \frac{f}{4\pi} \int \delta ({\bf y})
\frac{{\bf x}-{\bf y}}{\left| {\bf x}-{\bf y} \right|^3} d^3 {\bf
y}, \end{equation} where the growth index $f$ is defined by:
\begin{equation} f=\frac{d \ln \delta}{d \ln a},
 \label{eq5}
\end{equation}
and it is proportional to the ratio of the second term of equation
(\ref{eq3}) (friction) to the third (Poisson) term.

We use the evolution of the density contrast $\delta$ to compute the
growth index of structure $f$, which is an important quantity for
the interpretation of peculiar velocities of galaxies
\cite{P80,rah02}. Replacing the density contrast with the growth
index in equation (\ref{eq31}) results in the evolution of growth
index as:
\begin{eqnarray}
\label{index} &&\frac{df}{d\ln a}=
\frac{3H_0^2}{2\dot{a}^2a}\Omega_m
-f^2-f\left[1+\frac{\ddot{a}}{aH^2}\right].
\end{eqnarray}
The above equation in terms of dimensionless quantity is:
\begin{eqnarray}
\label{index1} &&\frac{df}{d\ln a}= \frac{3\Omega_m}{2a{\cal
H}^2(X)}-f^2-f\left[2+\frac{a{\cal H}'(X)}{{\cal
H}(X)}\frac{dX}{da}\right].
\end{eqnarray}

Figure (\ref{gi}) shows the numerical solution of equation
(\ref{index1}) in terms of redshift. As we expect increasing $X_0$
causes decreasing $\Omega_{m}$ from best fit and a decreasing in the
evolution of density contrast versus scale factor and growth index
in the small redshifts. This behavior is the same as what happens in
the $\Lambda$CDM model.

To put constraint on the model using large structure formation, we
rely to the observation of $220,000$ galaxies with the $2$dFGRS
experiment provides the numerical value of growth index
\cite{eisenstein05}. By measurements of two-point correlation
function, the $2$dFGRS team reported the redshift distortion
parameter of $\beta = f/b =0.49\pm0.09$ at $z=0.15$, where $b$ is
the bias parameter describing the difference in the distribution of
galaxies and their masses. Verde et al. (2003) used the bispectrum
of $2$dFGRS galaxies \cite{ver01,la02} and obtained $b_{verde}
=1.04\pm 0.11$ which gave $f= 0.51\pm0.10$. Now we fit the growth
index at the present time derived from the equation (\ref{index1})
with the observational value. This fitting gives a less constraint
to the parameters of the model, so in order to have a better
confinement of the parameters, we combine this fitting with those of
SNIa$+$CMB$+$SDSS which have been discussed in the previous section.
We perform the least square fitting by minimizing
$\chi^2=\chi_{\rm{SNIa}}^2+\chi_{\rm{CMB}}^2+\chi_{\rm{SDSS}}^2+\chi_{\rm{LSS}}^2$,
where
\begin{eqnarray}
\chi^2_{\rm
{LSS}}=\frac{[f_{obs}(z=0.15)-f_{th}(z=0.15;X_0)]^2}{\sigma_{f_{obs}}^2}
\end{eqnarray}
 The best fit value with the corresponding error bar for the
$X_0$ by using new Gold sample data is:
$X_0=6.192_{-0.177}^{+0.167}$ provides
$\Omega_m=0.278_{0.278}^{+0.273}$ at $1\sigma$ confidence level with
$\chi^2_{min}/N_{d.o.f}=0.900$. Using the SNLS supernova data, the
best fit value for model parameter is: $X_0=6.433_{-0.091}^{+0.089}$
gives $\Omega_m=0.252_{-0.150}^{+0.147}$ at $1\sigma$ confidence
level with $\chi^2_{min}/N_{d.o.f}=11.486$. The error bars have been
obtained through the likelihood functions $(\mathcal{L}$$\propto
e^{-\chi^2/2})$ marginalized over the nuisance parameter $h$. The
likelihood functions for the four cases of (i) fitting model with
Supernova data, (ii) combined analysis with the two experiments of
SNIa$+$CMB, iii) combined analysis with the three experiments of
SNIa$+$CMB$+$SDSS and (iv) combining all four experiments of
SNIa$+$CMB$+$SDSS$+$LSS are shown in Figures \ref{like1} and
\ref{like2}. The best fit value and age of universe computing in the
$f(R)$ modified gravity model are reported in table \ref{tab2}.

\begin{table}
\begin{center}
\caption{\label{tab2} The best values for the parameters of $f(R)$
modified gravity with the corresponding age for the universe from
fitting with SNIa from new Gold sample and SNLS data, SNIa+CMB,
SNIa+CMB+SDSS and SNIa+CMB+SDSS+LSS experiments at one and two
$\sigma$ confidence level. The value of $\Omega_m$ was determined
according to equation (\ref{matter})}
\begin{tabular}{|c|c|c|c|}
Observation & $X_0$&$\Omega_m$ & Age (Gyr)
\\ \hline
  &&& \\
 & $6.207^{+0.230}_{-0.147}$& $0.276_{-0.240}^{+0.376}$&$14.71 ^{+0.41}_{-0.23}$\\ 
 SNIa(new Gold)&&&\\
 & $6.207^{+0.445}_{-0.515}$&$0.276_{-0.276}^{+0.727}$ &$14.71 ^{+0.85}_{-0.75}$
\\ &&&\\ \hline
&&&\\
SNIa(new Gold)+& $6.190^{+0.224}_{-0.242}$&$0.278_{-0.278}^{+0.365}$ &$14.69^{+0.39}_{-0.37}$ \\
&&&\\
CMB &$6.190^{+0.433}_{-0.503}$&$0.278_{-0.278}^{+0.706}$&$14.69^{+0.82}_{-0.72}$ \\
 &&&\\ \hline
 &&&\\
SNIa(new Gold)+& $6.169^{+0.170}_{-0.184}$&$0.281_{-0.281}^{+0.277}$&$14.56^{+0.29}_{-0.28}$ \\
&&&\\
CMB+SDSS &$6.169^{+0.327}_{-0.380}$&$0.281_{-0.281}^{+0.533}$& $14.56^{+0.58}_{-0.56}$\\
&&&\\ \hline
&&& \\
SNIa(new Gold)+&
$6.192^{+0.167}_{-0.177}$&$0.278_{-0.278}^{+0.273}$&$14.69^{+0.29}_{-0.28}$
 \\
 &&&\\
 CMB+SDSS+LSS&$6.192^{+0.321}_{-0.368}$&$0.278_{-0.278}^{+0.524}$& $14.69^{+0.55}_{-0.58}$ \\
&&&\\  
 \hline &&&\\

  & $6.484^{+0.099}_{-0.099}$&$0.246_{-0.164}^{+0.164}$ &$13.69^{+0.18}_{-0.17}$ \\ 
 SNIa (SNLS)&&&\\
 & $6.484^{+0.196}_{-0.201}$&$0.246_{-0.246}^{+0.324}$&$13.69^{+0.37}_{-0.35}$
\\ &&&\\ \hline
&&&\\
SNIa(SNLS)+& $6.477^{+0.098}_{-0.100}$&$0.246_{-0.165}^{+0.162}$ &$13.68^{+0.18}_{-0.17}$ \\
&&&\\
CMB &$6.477^{+0.194}_{-0.200}$&$0.246_{-0.246}^{+0.321}$&$13.68^{+0.36}_{-0.33}$ \\
&&&\\ \hline
&&&\\
SNIa(SNLS)+& $6.428^{+0.090}_{-0.092}$&$0.252_{-0.152}^{+0.148}$&$13.60^{+0.16}_{-0.15}$ \\
&&&\\
CMB+SDSS &$6.428^{+0.177}_{-0.185}$&$0.252_{-0.252}^{+0.292}$&$13.60^{+0.32}_{-0.30}$ \\
&&&\\ \hline
&&& \\
SNIa(SNLS)+&
$6.433^{+0.089}_{-0.091}$&$0.252_{-0.150}^{+0.147}$&$13.61^{+0.16}_{-0.15}$\\
&&& \\
 CMB+SDSS+LSS&$6.433^{+0.176}_{-0.183}$&$0.252_{-0.252}^{-0.291}$& $13.61^{+0.32}_{-0.30}$ \\
&&&\\
\end{tabular}
\end{center}
\end{table}

\begin{table}[htp]
\begin{center}
\caption{\label{tab3} The value of $\tau$ for three high redshift
objects, using the parameters of the model derived from fitting with
the observations at one and two $\sigma$ level of confidences.}
\begin{tabular}{|c|c|c|c|}
  Observation & LBDS &LBDS  & APM  \\
& $53$W$069$&$53$W$091$& $08279+5255$ \\
  & $z=1.43$&$z=1.55$& $z=3.91$  \\ \hline
  &&&\\
SNIa (new Gold)& $1.23^{+0.05}_{-0.03}$ & $1.32^{+0.05}_{-0.03}$& $0.86^{+0.37}_{-0.05}$ \\
&&&\\
&$1.23^{+0.10}_{-0.09}$ & $1.32^{+0.11}_{-0.10}$&$0.86^{+0.38}_{-0.08}$\\
&&&\\
 \hline

&&&\\
SNIa(new Gold)+CMB &$1.23^{+0.05}_{-0.04}$&$1.32^{+0.05}_{-0.05}$&$0.86^{+0.37}_{-0.06}$ \\
 & && \\

&$1.23^{+0.10}_{-0.08}$&$1.32^{+0.11}_{-0.09}$&$0.86^{+0.38}_{-0.09}$\\
&&& \\\hline

&&&\\
SNIa(new Gold)+CMB &$1.23^{+0.03}_{-0.04}$&$1.32^{+0.04}_{-0.04}$&$0.85^{+0.36}_{-0.05}$ \\
 +SDSS& && \\
&$1.23^{+0.07}_{-0.06}$&$1.32^{+0.08}_{-0.07}$&$0.85^{+0.37}_{-0.07}$\\
&&&\\\hline
&&&\\
SNIa(new Gold)+CMB & $1.23^{+0.04}_{-0.03}$&$1.32^{+0.04}_{-0.04}$&$0.86^{+0.36}_{-0.05}$ \\
 +SDSS+LSS& && \\
 & $1.23^{+0.07}_{-0.06}$&$1.32^{+0.08}_{-0.08}$&$0.86^{+0.37}_{-0.07}$\\
 &&&\\ \hline
&&&\\
 SNIa (SNLS)& $ 1.16^{+0.02}_{-0.02}$ & $1.21^{+0.02}_{-0.02}$& $0.82^{+0.35}_{-0.04}$ \\ &&&\\
 &$ 1.16^{+0.02}_{-0.02}$ & $1.21^{+0.02}_{-0.02}$& $0.82^{+0.35}_{-0.04}$ \\
 &&&\\ \hline
&&&\\
SNIa(SNLS)+CMB & $1.16^{+0.02}_{-0.02}$&$1.25^{+0.02}_{-0.02}$&$0.82^{+0.35}_{-0.02}$ \\
 & && \\
&
$1.16^{+0.05}_{-0.04}$&$1.25^{+0.05}_{-0.04}$&$0.82^{+0.36}_{-0.06}$\\
&&&\\
 \hline
&&&\\
SNIa(SNLS)+CMB & $1.15^{+0.02}_{-0.02}$&$1.24^{+0.03}_{-0.02}$&$0.81^{+0.35}_{-0.04}$ \\
 +SDSS& && \\
&
$1.15^{+0.04}_{-0.04}$&$1.24^{+0.04}_{-0.04}$&$0.81^{+0.36}_{-0.05}$\\
&&&\\\hline
&&&\\
SNIa(SNLS)+CMB & $1.15^{+0.02}_{-0.02}$&$1.24^{+0.02}_{-0.02}$&$0.81^{+0.35}_{-0.04}$ \\
 +SDSS+LSS& && \\
&
$1.15^{+0.04}_{-0.03}$&$1.24^{+0.04}_{-0.04}$&$0.81^{+0.36}_{-0.05}$\\
&&&\\
\end{tabular}
\end{center}
\end{table}
\section{Age of Universe}
\label{sec6}
 The age of universe integrated from the big bang up to
now for flat universe in terms of $X_0$ is given by:
\begin{eqnarray}\label{age55}
t_0(X_0) &=& \int_0^{t_0}\,dt =\int_0^\infty {dz\over (1+z) H(z)}\nonumber\\
&=&\frac{1}{3H_0}\int_{X_0}^{\infty}\frac{F'-XF''}{2F-XF'}\frac{dX}{{\cal{H}}(X)}
\end{eqnarray}

Figure~\ref{age} shows the dependence of $H_0t_0$ (Hubble parameter
times the age of universe) on $X_0$ for a flat universe. In the
lower panel we show it for $\Lambda$CDM for comparison. As we expect
modified gravity behaves as a dark energy and increasing $X_0$
($\Omega_{\lambda}$) result in a longer age for the universe in the
$f(R)$ modified gravity model.

 The "age crisis" is one the main reasons
of the acceleration phase of the universe. The problem is that the
universe's age in the Cold Dark Matter (CDM) universe is less than
the age of old stars in it. Studies on the old stars
\cite{carretta00} suggest an age of $13^{+4}_{-2}$ Gyr for the
universe. Richer et. al. \cite{richer02} and Hasen et. al.
\cite{hansen02} also proposed an age of $12.7\pm0.7$ Gyr, using the
white dwarf cooling sequence method (for full review of the cosmic
age see \cite{spe03}). To do another consistency test, we compare
the age of universe derived from this model with the age of old
stars and Old High Redshift Galaxies (OHRG) in various redshifts.
Table \ref{tab2} shows that the age of universe from the combined
analysis of SNIa$+$CMB$+$SDSS$+$LSS is $14.69_{-0.28}^{+0.29}$ Gyr
and $13.61_{-0.15}^{+0.16}$ Gyr for new Gold sample and SNLS data,
respectively. These values are in agreement with the age of old
stars \cite{carretta00}. Here we take three OHRG for comparison with
the power-law dark energy model, namely the LBDS $53$W$091$, a
$3.5$-Gyr old radio galaxy at $z=1.55$ \cite{dunlop96}, the LBDS
$53$W$069$ a $4.0$-Gyr old radio galaxy at $z=1.43$ \cite{dunlop99}
and a quasar, APM $08279+5255$ at $z=3.91$ with an age of
$t=2.1_{-0.1}^{+0.9}$Gyr \cite{hasinger02}. The latter has once
again led to the "age crisis". An interesting point about this
quasar is that it cannot be accommodated in the $\Lambda$CDM model
\cite{jan06}. To quantify the age-consistency test we introduce the
expression $\tau$ as:
\begin{equation}
 \tau=\frac{t(z;X_0)}{t_{obs}} = \frac{t(z;X_0)H_0}{t_{obs}H_0},
\end{equation}
where $t(z)$ is the age of universe, obtained from the equation
(\ref{age55}) and $t_{obs}$ is an estimation for the age of old
cosmological object. In order to have a compatible age for the
universe we should have $\tau>1$. Table \ref{tab3} reports the value
of $\tau$ for three mentioned OHRG with various observations. We see
that $f(R)$ modified gravity with the parameters from the combined
observations, provides a compatible age for the universe, compared
to the age of old objects, while  the SNLS data result in a shorter
age for the universe. Once again, APM $08279+5255$ at $z=3.91$ has a
longer age than the universe but gives better results than most
cosmological models  \cite{sa1,sa2}.
\begin{figure}[t]
\epsfxsize=9.5truecm\epsfbox{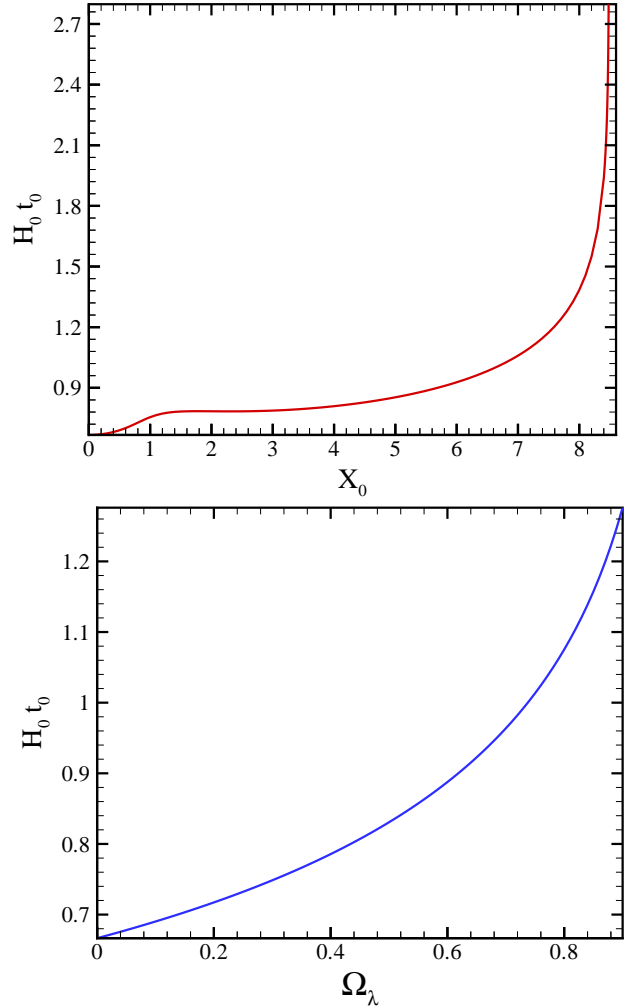} \narrowtext \caption{
$H_0t_0$ (age of universe times the Hubble constant at the present
time) as a function of $X_0=R_0/H_0^2$ (upper panel). $H_0t_0$ for
$\Lambda$CDM versus $\Omega_{ \lambda}$ (lower panel). Increasing
$X_0$ gives a longer age for the universe. This behavior is the same
as what happens in $\Lambda$CDM.} \label{age}
\end{figure}

\section{conclusion}
\label{conc} Here in this work we obtained the dynamics of universe
with $f(R)=\sqrt{R^2-R_0^2}$ gravity and compared it with recent
cosmological observations. This comparison has been performed with
Supernova Type Ia Gold sample and SNLS supernova data, CMB shift
parameter, location of baryonic acoustic oscillation peak observed
by SDSS and large scale structure formation data by $2$dFGRS. The
best parameters obtained from fitting with the new Gold sample data
are: $h=0.63$, $X_0=6.192_{-0.177}^{+0.167}$ provides
$\Omega_m=0.278_{-0.278}^{+0.273}$ at $1\sigma$ confidence level
with $\chi^2_{min}/N_{d.o.f}=0.900$. Using the SNLS supernova data,
the best fit value for model parameter is:
$X_0=6.433_{-0.091}^{+0.089}$ gives
$\Omega_m=0.252_{-0.150}^{+0.147}$ at $1\sigma$ confidence level
with $\chi^2_{min}/N_{d.o.f}=11.486$. We also performed the age
test, comparing the age of old stars and old high redshift galaxies
with the age derived from this model. From the best fit parameters
of the model using new Gold sample and SNLS, we obtained an age of
$14.69_{-0.28}^{+0.29}$ Gyr and $13.61_{-0.15}^{+0.16}$ Gyr,
respectively, for the universe which is in agreement with the age of
old stars. We also chose two high redshift radio galaxies at
$z=1.55$ and $z=1.43$ with a quasar at $z=3.91$. The ages of the two
first objects were consistent with the age of universe, means that
they were younger than the universe while the third one was not.


\end{document}